\documentclass[12pt]{article}
\usepackage{a4}
\usepackage{epsf}
\usepackage{amssymb}
\usepackage{cite}
\def\Bbb{\mathbb}
\def\BZ{\Bbb Z} \def\BR{\Bbb R}
\def\BC{\Bbb C} \def\BP{\Bbb P}

\catcode`@=11 \@addtoreset{equation}{section} \catcode`@=12

\begin{document}
\begin{titlepage}
\noindent
{\tt TIFR/TH/01-22} \hfill
{\tt hep-th/0108234}\\
{\tt IMSc/2001/07/32}~~~
\hfill August, 2001
\vfill
\begin{center}
{\Large \bf Disc Instantons in Linear Sigma Models} \\[1cm]
Suresh Govindarajan\footnote{Email: suresh@chaos.iitm.ernet.in}\\
{\em Department of Physics, Indian Institute of Technology, Madras,\\
Chennai 600 036, India\\[10pt]}
T. Jayaraman\footnote{Email: jayaram@imsc.ernet.in}\\
{\em The Institute of Mathematical Sciences, \\ Chennai 600 113, India\\
[10pt]}
Tapobrata Sarkar\footnote{Email: tapo@theory.tifr.res.in}\\
{\em Department of Theoretical Physics, \\ Tata Institute of 
Fundamental Research, \\ Homi Bhabha Road, Mumbai 400 005, India}\\ 
\end{center}
\vfill
\begin{abstract}
We construct a linear sigma model for open-strings ending on
special Lagrangian cycles of a Calabi-Yau manifold. 
We illustrate the construction
for the cases considered by Aganagic and Vafa(AV). This leads naturally to
concrete models for the  moduli space of open-string instantons.
These instanton moduli spaces can be seen to be intimately related to certain
auxiliary boundary toric varieties. By considering the relevant 
Gelfand-Kapranov-Zelevinsky (GKZ) differential
equations of the boundary toric variety, we obtain the contributions to the
worldvolume superpotential on the A-branes from open-string instantons.
By using an ansatz due to Aganagic, Klemm and Vafa (AKV), we obtain the 
relevant
change of variables from the linear sigma model to the non-linear sigma model
variables - the open-string mirror map. Using this
mirror map, we obtain results in agreement  with those of AV and AKV
for the counting of holomorphic disc instantons.
\end{abstract}
\vfill
\end{titlepage}

\section{Introduction}

D-branes have come to play a fundamental role 
in our understanding of string theory, especially for
the case Calabi-Yau (CY) compactifications. 
Subsequent to  the initial 
description of D-branes wrapping supersymmetric cycles of CY manifolds 
\cite{ooy}, it was realised that A and B-type 
D-branes\footnote{The terminology 
arising from the topological theory that support these branes, in accordance
with the open-string world sheet boundary conditions that preserve 
some amount of supersymmetry in either case. We shall refer to the
D-branes as A-branes/B-branes indicating the appropriate
boundary condition.} are closely related to 
the objects that are relevant to Kontsevich's homological mirror symmetry 
proposal\cite{kontsevich,vafa}.  
The fact that mirror symmetry, generally speaking,
exchanges A-branes and B-branes 
has begun to play a crucial role in understanding several aspects of 
stringy geometry on Calabi-Yau manifolds. 

One of the applications of mirror symmetry in type II compactifications
has been the computation of closed string instanton corrections
to various physical quantities. With the inclusion of D-brane sectors
for strings moving on Calabi-Yau backgrounds,
one has, in addition, to consider a generalisation
of mirror symmetry that also includes the open string sectors. In
particular, one may expect that mirror symmetry in the open and closed
string sectors taken together is relevant to understanding the
contribution of open-string instanton corrections.

In contrast to the case of purely closed strings on CY, it has been
pointed out that various quantities in the world-volume theories of both
A and B type D-branes acquire open-string instanton corrections. In
general it is expected that the F-terms in the world-volume theory have
open-string instanton corrections for A-branes whereas the D-terms are
the ones that acquire instanton corrections for
B-branes\cite{quintic,douglasstrings}. Various
general considerations relevant to open-string instanton corrections in
the case of compact Calabi-Yau manifolds as well as non-compact
Calabi-Yau manifolds have been studied earlier\cite{kachruetal,lazar}.

However the explicit calculation of open-string instanton corrections 
to the world-volume superpotential for A-branes
using mirror symmetry have first appeared recently in the important work of
Aganagic and Vafa \cite{AV} and Aganagic, Klemm and Vafa\cite{AKV}. 
The authors in these papers have computed
the world-volume superpotential for some non-compact 
A-branes in non-compact Calabi-Yau manifolds (in particular, those that
can be described without a world-sheet superpotential), 
by appealing to mirror symmetry and using the holomorphic Chern-Simons
action on the B-brane side to reduce the problem to a more tractable 
computation. This computation uses and extends the techniques first described
in \cite{hv,hiv}, whereby  the mirror transformation, in the
presence of D-branes, can be explicitly implemented as a world-sheet 
duality transformation involving the 
fields of a Gauged Linear Sigma Model (GLSM) that describes these
D-brane configurations.  
Remarkably, the authors of \cite{AV,AKV} showed that the
(instanton generated) A-brane superpotential thus computed conformed 
exactly to the strong integrality predictions of such quantities from 
completely different considerations of large $N$ duality \cite{ov}.

We briefly recall here the manner in which this integrality
predictions were first made.
In \cite{gv1} it was conjectured that
the large $N$ limit of Chern-Simons theory on $S^3$ is dual
to A-type closed topological string theory on the resolved conifold, with 
a particular identification between the parameters of the two theories.
This conjecture was verified 
at the level of the partition function on both sides. The conjecture 
was extended to the observables of the 
Chern-Simons theory, namely the knot invariants in \cite{ov}. By identifying
the extra D-branes (associated with the knots on the Chern-Simons side)
on the resolved conifold, non-trivial predictions were made in 
\cite{ov} regarding the structure of the topological string amplitudes, 
using ideas of \cite{gv2}. (Knot theoretic aspects of the conjecture 
were verified in \cite{knots}.)
Further, from the fact that topological
disc amplitudes compute superpotential terms in $N=1$ $d=4$ gauge
theories, a general structure for the A-type superpotential was proposed,
which had the integrality properties that we referred to above.

It is of interest to ask the question whether the A-brane world-volume
superpotential, together with the open-string instanton corrections, can
be directly computed without explicit recourse to mirror symmetry,
particularly in the framework of the GLSM\cite{wittenphases}. 
In this paper,
we will take the first steps in this direction. We will show that the 
description of the
A-branes in the linear sigma model framework suggests a natural and
concrete description of  these (partially compactified) open-string 
instanton moduli 
spaces. These instanton moduli spaces will be seen to be intimately
related to some auxiliary toric varieties that we shall refer to as 
{\em boundary toric varieties}. For these boundary toric varieties we will
write down the appropriate  
Gelfand-Kapranov-Zelevinsky (GKZ) differential equations, by an extension
of the methods developed for the closed string case, particularly for
non-compact CY manifolds  by \cite{yauetal,chiangetal}. We will then show how
the relevant solutions of the GKZ equations describe the world-volume
superpotential for the A-branes in the examples considered by
\cite{AV,AKV}. We will also show how to obtain the analogue of the
 open-string mirror map in this context, using in part, an ansatz due to 
Aganagic, Klemm and Vafa.

This paper is organised as follows. In section 2, we describe some basic
facts about the Gauged Linear Sigma Model. After describing the 
construction of A and B-type superspace, we discuss the boundary
conditions of \cite{AV} that describes A-branes. In section 3, we
discuss the GLSM for A-branes and the boundary conditions therein. 
In section 4, we discuss the topological aspects of the A-model,
before moving on, in section 5, to the open string instanton moduli
space. After discussing briefly certain issues of 
stability, we introduce the boundary toric variety. In section 6,
we consider aspects of local mirror symmetry and the GKZ system of
equations for our boundary toric variety. We discuss how to obtain the
bulk and boundary periods that are relevant for the computation of the
A-brane superpotential in the mirror. Section 7 deals with some examples
in order to illustrate the methods. We conclude with some observations
in section 8.

\section{Background}

The supersymmetry generators for $(2,2)$ supersymmetry
can be written as (in the notation of \cite{wittenphases})
\begin{eqnarray}
\delta_{susy} &=& \epsilon^\alpha Q_\alpha + \bar{Q}_{\dot{\alpha}} 
\epsilon^{\dot{\alpha}} \nonumber \\
&=& -\epsilon_- Q_+ + \epsilon_+ Q_- + \bar{\epsilon}_- \bar{Q}_+
- \bar{\epsilon}_+ \bar{Q}_-
\end{eqnarray}
In superspace with coordinates $(x^m,\theta^\pm,\bar{\theta}^\pm)$, 
the supersymmetry generators have the following representation
\begin{eqnarray}
Q_\pm&=&{\partial\over{\partial \theta^\pm}}+ i \bar{\theta}^\pm (\partial_0\pm
\partial_1) \nonumber \\
\bar{Q}_\pm &=& -{\partial\over{\partial \bar{\theta}^\pm} }
- i \theta^\pm (\partial_0 \pm \partial_1) 
\end{eqnarray}

On the boundary, one set of linear combinations
of this supersymmetry is preserved. Let us denote the unbroken
supersymmetry generators by $Q$ and $\bar{Q}$.  Let us denote the
boundary in superspace by the coordinates $(x^0,\theta,\bar{\theta})$.
Then, the supersymmetry generators can be given the following representation
\begin{equation}
Q= {\partial\over{\partial\theta}} + i \overline{\theta}
\partial_0 \quad,\quad
\overline{Q}= -{\partial\over{\partial\overline{\theta}}} - i \theta
\partial_0 \quad,
\end{equation}
We will now work out the precise relationship between the boundary coordinates/
supersymmetry generators. We will deal with the A- and B-type cases separately.

\subsection*{\normalsize A-type superspace}
Under A-type supersymmetry, let 
$$
\epsilon_A\equiv -\sqrt2\epsilon_- = -\sqrt2\eta \bar{\epsilon}_+\quad,
$$
where we use the subscript $A$ to denote
that we are considering boundaries that preserve A-type supersymmetry.
Then, on using
\begin{eqnarray}
\delta_A 
&\equiv& \epsilon_A Q_A - \bar{\epsilon}_A \bar{Q}_A \nonumber \\
&=& - \bar{\epsilon}_+ (\bar{Q}_- + \eta Q_+)
           + \epsilon_+ (Q_- + \eta \bar{Q}_+)\quad, 
\end{eqnarray}
we obtain
$$ Q_A = \left({{Q_+ + \eta \bar{Q}_-}\over \sqrt2}\right) \quad  
{\rm and} \quad
\theta_A\equiv \sqrt2 \theta^+ = -\sqrt2 \eta \bar{\theta}^- \quad.
$$

\subsection*{\normalsize B-type superspace}
In a similar fashion one chooses $\epsilon_B \equiv \sqrt2 
\epsilon_- =\sqrt2 \eta \epsilon_+$. 
\begin{eqnarray}
\delta_B &=& \epsilon_B \left({{-Q_+ + \eta Q_-}\over\sqrt2} \right)
- \bar{\epsilon}_B \left({{-\bar{Q}_+ + \eta \bar{Q}_-}\over\sqrt2}\right)
\nonumber \\
&\equiv& \epsilon_B Q_B - \bar{\epsilon}_B \bar{Q}_B \quad,
\end{eqnarray}
from which we identify 
$$
Q_B = {{-Q_+ + \eta Q_-}\over\sqrt2}\quad{\rm and}\quad
\theta_B\equiv -\sqrt2 \theta^+ = \sqrt2 \eta \theta^-\quad.$$

\subsection{Boundary multiplets}

In boundary superspace with coordinates $(x^0,\theta,\bar{\theta})$,
the superderivatives are
\begin{equation}
D= {\partial\over{\partial\theta}} - i \overline{\theta}
\partial_0 \quad,\quad
\bar{D}= -{\partial\over{\partial\overline{\theta}}} + i \theta
\partial_0 \quad.
\end{equation}
We will deal with two kinds of boundary multiplets: a boundary
chiral multiplet and an unconstrained complex multiplet.

A {\bf boundary chiral multiplet} satisfying the constraint
$$
\bar{D} L = 0 \quad,
$$ 
which has the following component expansion
$$
L = l + i \theta \lambda - i \theta \bar{\theta} \partial_0 l\quad.
$$

An unconstrained {\bf complex superfield} $U$ with component expansion
\begin{equation}
U = u + \theta \upsilon - \bar{\theta} \xi +\theta \bar{\theta}  (D'+iR)
\end{equation}
where $D'$ and $R$ are real fields.
Under supersymmetry, the transformations are
\begin{eqnarray}
\delta u &=& -\bar{\epsilon} \xi + \epsilon \upsilon \nonumber \\
\delta \upsilon &=& -\bar{\epsilon} (-i\partial_0 u +D'+ iR) \\
\delta \xi &=& -\epsilon(i\partial_0u +D'+iR) \nonumber \\
\delta (D'+iR) &=& -i \epsilon \partial_0 \upsilon 
- i \bar{\epsilon} \partial_0 \xi
\nonumber
\end{eqnarray}

Now consider a complex  superfield with the following gauge invariance
\begin{equation}
U \rightarrow U + L
\end{equation}
where $L$ is a chiral superfield. 
A gauge-invariant observable is
given by $\Xi\equiv\bar{D} U$. 
The invariance trivially follows from the chirality of $L$. 
$\Xi$ has the following superfield expansion
\begin{equation}
\Xi = \bar{D} U = \xi + \theta (D'+i R + i\partial_0 u) -i \theta\bar{\theta} 
\partial_0 \xi\quad,
\end{equation}

We will identify the real part of
$U$ with the bosonic $u$ discussed earlier. Keeping this in mind,
we can partially gauge-fix the imaginary part of $U$ as follows
\begin{equation}
{\rm Im}(U) = \theta \bar{\theta}  (R)\quad.
\end{equation}
Note that the gauge choice is algebraic and hence we will call
it the Wess-Zumino (WZ) gauge.
This leaves us with the following gauge invariance which acts as
\begin{eqnarray}
\delta {\rm Re}(u) = {\rm Re} (l) \nonumber  \\
\delta R = -\partial_0 {\rm Re} (l)
\end{eqnarray}
In the WZ gauge,
the field $U$ has the following superfield expansion
\begin{equation}
U = u + \theta \bar{\xi} - \bar{\theta} \xi +\theta \bar{\theta} (D'+iR) 
\end{equation}
where $u$ is now a  real field. Thus, if $R=0$, it would
be a real superfield.

The supersymmetry and gauge
transformations which preserve the WZ gauge are
\begin{eqnarray}
\delta u &=&  -\bar{\epsilon} \xi + \epsilon \bar{\xi}+ l \nonumber \\
\delta \xi &=& -\epsilon(i\partial_0u +D'+iR)  \\
\delta D' &=&  -i \epsilon \partial_0\bar{\xi}
- i\bar{\epsilon} \partial_0 \xi \nonumber \\
\delta  R &=& - \partial_0 l \nonumber 
\end{eqnarray}
where $l$ is now a real parameter. Notice that $R$ is invariant under
supersymmetry and transforms as $-\partial_0 l$ under gauge transformations
and thus behaves as a gauge field.

\subsection{Decomposition of bulk multiplets: A-type}

First, consider a chiral superfield $\Phi$ which has the following component
expansion:
\begin{eqnarray}
\Phi &=& \phi + \sqrt2 \theta^+ \psi_+ + \sqrt2 \theta^- \psi_- 
+ 2\theta^+\theta^- F  
- i \theta^+\bar{\theta}^+ (D_0 + D_1)\phi\nonumber \\
&-& i \theta^-\bar{\theta}^- (D_0 - D_1)\phi
+ i \theta^-\bar{\theta}^+ \sqrt2 Q \sigma \phi
- i \theta^+\bar{\theta}^- \sqrt2 Q \bar{\sigma} \phi \nonumber \\
&-& i \sqrt2 \theta^+\theta^-\bar{\theta}^- [ (D_0 - D_1)\psi_+ 
+ \sqrt2 Q \bar{\sigma} \psi_-] \nonumber \\
&+& i \sqrt2 \theta^+\theta^-\bar{\theta}^+ [(D_0 + D_1)\psi_- 
+ \sqrt2 Q \sigma \psi_+]
\end{eqnarray}
In order to obtain
the A-type superfield, we need to set $\theta\equiv \sqrt2\theta^+ 
= -\sqrt2\eta \bar{\theta}^-$
Thus, we obtain
\begin{equation}
\Phi' \equiv \Phi|_{\theta^+ = -\eta \bar{\theta}^-}
= \phi + \theta \psi_+ - \eta \bar{\theta} \psi_- -
\theta \bar{\theta} (\eta F +i D_1\phi)
\end{equation}
where the prime indicates an A-type superfield. This is a complex unconstrained 
superfield.

Next, let us consider a twisted chiral superfield $\Sigma$ which has
the following component expansion
\begin{eqnarray}
\Sigma &=& \sigma + i \sqrt 2\theta^+
\bar{\lambda}_+ -i\sqrt 2\bar{\theta}^-\lambda_- +
\sqrt 2\theta^+\bar{\theta}^-(D-iv_{01}) \nonumber \\
&-&i\bar{\theta}^-\theta^-(\partial_0
-\partial_1)\sigma-i\theta^+\bar{\theta}^+(\partial_0+\partial_1)\sigma
+\sqrt 2\bar{\theta}^-\theta^+\theta^-(\partial_0-\partial_1)\bar{\lambda}_+
\nonumber \\
&+&\sqrt 2\theta^+\bar{\theta}^-\bar{\theta}^+(\partial_0+\partial_1)\lambda_-
-\theta^+\bar{\theta}^-\theta^-\bar{\theta}^+
(\partial_0{}^2-\partial_1{}^2)\sigma.
\end{eqnarray}
We define the A-type superfield $\Sigma'$ as in the case of the chiral 
superfield. We obtain
\begin{equation}
\Sigma' \equiv \Sigma|_{\theta^+ = -\eta \bar{\theta}^-}
= \sigma +i \theta (\bar{\lambda}_+ + \eta \lambda_-)
-i\theta \bar{\theta} \partial_0 \sigma 
\end{equation}
which is the superfield expansion for an A-type chiral superfield. In addition,
one obtains a Fermi chiral superfield $\Lambda$ whose lowest component is 
$\lambda\equiv (\bar{\lambda}_+ - \eta \lambda_-)$ satisfying
\begin{equation}
\bar{D} \Lambda = 2 \partial_1 \Sigma'\quad,
\end{equation}
where one treats $\partial_1\Sigma'$ as an A-type chiral superfield with
lowest component $\partial_1 \sigma$. The component expansion for $\Lambda$ 
is
\begin{equation}
\Lambda = (\bar{\lambda}_+ - \eta \lambda_-) + i\sqrt2 \theta (D-iv_{01})
-2 \overline{\theta} \partial_1\sigma 
+ \theta \overline{\theta} \left[ 
-i\partial_0 (\bar{\lambda}_+ - \eta \lambda_-) 
+2i \partial_1 (\bar{\lambda}_+ + \eta \lambda_-)\right]
\end{equation}
Something interesting happens if $D-iv_{01}$ is set to zero on the boundary
(part of which we know is required by the vanishing of ordinary variations).
Then, using the bulk equations of motion for the $\lambda$'s, one can
see that $\Lambda$ becomes an {\em anti-chiral} boundary superfield.

\subsection{The AV boundary conditions}

Aganagic and Vafa\cite{AV} have constructed a class of special Lagrangian(sL)
cycles in toric varieties by adapting the construction of
Harvey and Lawson in $\BC^n$.
We shall describe this construction in the language of the GLSM.
Let $\Phi_i$ ($i=1,\ldots,n$)
be the chiral superfields with $U(1)$ charges $Q_i^a$
which appear in the gauged linear sigma model associated with 
the toric variety.
The toric variety is obtained by imposing
the (bulk) D-term constraint given by
\begin{equation}
{D^a\over{e^2}} = r^a - \sum_i Q^a_i |\phi_i|^2 =0
\end{equation}
where $a=1,\ldots,h_{1,1}$. After taking into account
the gauge invariances involved,
one ends up with some weighted projective space (actually, a line
bundle on that space) of dimension $(n-h_{1,1})$. This is
the standard procedure called {\em symplectic reduction}.
The Calabi-Yau condition is given by $\sum_i Q_i^a=0$.

Lagrangian submanifolds in $\BC^n$
are specified by the following D-term like
conditions
\begin{equation}
\sum_i q_i^\alpha |\phi_i|^2 = c^\alpha\qquad \alpha=1,\ldots,r
\end{equation}
Note however that there is no associated local $U(1)$ invariance 
as occurs in symplectic reduction.
One further imposes $(n-r)$ additional conditions on linear
combinations of the phases ${\rm Im}\log \phi_i$
of the fields $\phi_i$ 
\begin{equation}
\sum_i v^\beta_i\  {\rm Im}\log \phi_i=0 \quad \beta=1,\ldots,(n-r)
\end{equation}
The $(q^\alpha,v^\beta)$ specify a Lagrangian submanifold of
$\BC^n$ (with the standard symplectic structure)
if 
$$\sum_i q^\alpha_i v^\beta_i =0\quad.$$
In order for this Lagrangian submanifold in $\BC^n$
to descend to the toric variety,
we choose $h_{1,1}$ of the conditions given above
to be the D-term conditions associated with the toric variety.
The conditions $\sum_i Q^a_iv^\beta_i=0$
imply that the conditions on the
phases are invariant under $U(1)$ gauge transformations.
Further, it follows that the
submanifold is special Lagrangian if $$\sum_i q^\alpha_i =0\quad.$$
This is clearly compatible with the Calabi-Yau condition on the bulk
$U(1)$ charges. 

For generic values of $c^\alpha$, the sL submanifolds
have a boundary. One can obtain sL submanifolds without boundary 
by a  doubling procedure. However, there exist non-generic situations
(such as the case when one of the $c^\alpha$ vanish)
where the doubling is not necessary. In other words, when we take
an sL submanifold without boundary for generic values of $c^\alpha$
to the non-generic value, the sL manifold splits into 
two\footnote{or a suitable
power of two depending on the number of doublings involved}. We shall
refer to one component as an {\em half-brane} following AV\cite{AV}.
In the toric description, {\em half-branes} lie on the edges of toric
skeletons.

In order to be more concrete, let us consider a specific example
furnished by the non-compact Calabi-Yau manifold given by the total 
space of the line-bundle ${\cal O}(-3)$ over $\BP^2$.
This non-compact Calabi-Yau manifold is described in the GLSM by
four fields with $U(1)$ charge $Q=(-3,1,1,1)$ and the boundary
conditions of the special Lagrangian submanifold are given by
$q^1=(-1,1,0,0)$, $q^2=(-1,0,1,0)$ and $v=(1,1,1,1)$. For generic
values of $c^\alpha$, the topology of the sL submanifolds (after
doubling) is $\BR \times S^1 \times S^1$. However, for suitable
choices of $c^\alpha$ (see below), one obtains half-branes which are
topologically $\BC\times S^1$,

Following the discussion  in AKV\cite{AKV}, there are three phases
(half-branes) of
interest: Phase I given by $c^2=0$ and $0<c^1 < r$,
Phase II given by $c^1=0$ and $0<c^2 < r$ and Phase III is
given by $c^1=c^2$ and $c^1<0$. The relevant phase diagram is
as follows
\begin{figure}[ht]
\begin{center}
\leavevmode\epsfysize=7cm \epsfbox{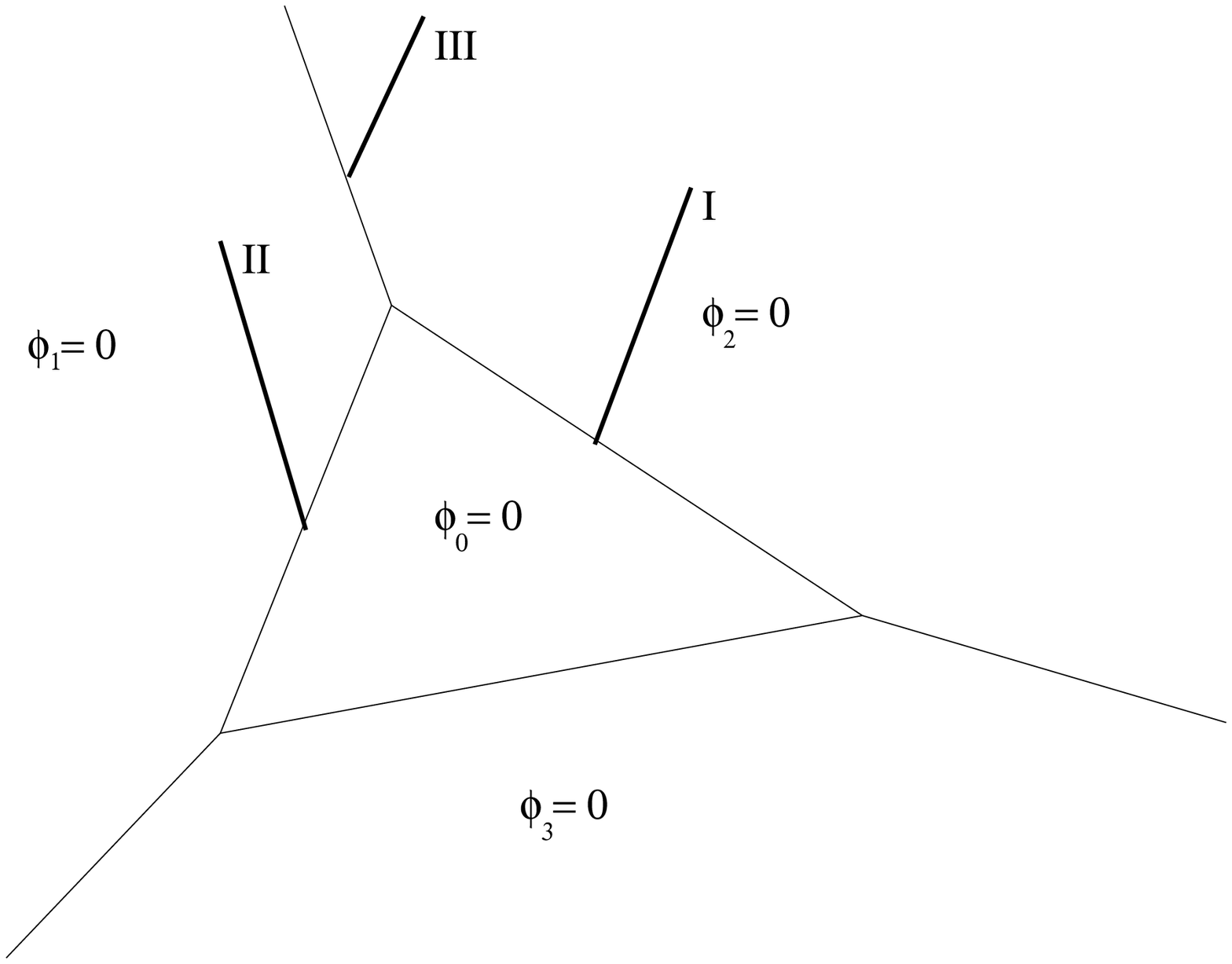}
\end{center}
\caption{Phases (half-branes) for ${\cal O}(-3)$ on $\BP^2$}
\end{figure}

In phase I, for large values of $c^1$ and $r$ when quantum corrections
are small, $c^1$ measures the size of a holomorphic disc ending on
the half-brane. This is the real part of a complex modulus, $\widehat{w}^1$,
whose imaginary part corresponds to turning on a Wilson line
wrapping the $S^1$ of the half-brane. $\widehat{w}^2$ is the other modulus
associated with the other $S^1$ which appears for generic values of
$c^\alpha$ vanishes on the half-brane. These values suffer quantum
corrections which become relevant for small values of $c^1$ and $r$.
We will denote the quantum corrected quantities by unhatted quantities.
However, $w^2$ will not be independent and will be determined as
a function of $w^1$ on the half-brane.

\subsection{The AV/AKV computation of the A-brane Superpotential}

As we mentioned in the introduction, the essential idea
of \cite{AV,AKV} is to compute the superpotential for 
the A-branes described in the last subsection  
 by using mirror symmetry to calculate
this in the B-model. Here, we will briefly review their arguments. 
We will be interested in the case of a space-filling
D6-brane wrapping a sL cycle of a non-compact CY three-fold.

It was shown in \cite{ov} that the amplitude of interest, the 
(instanton generated) disc amplitude in the A-model, is of the form
\begin{equation}
W=\sum_{n=0}^{\infty}\sum_{{\vec m},{\vec k}}\frac{
d_{{\vec m},{\vec k}}}{n^2}q^{n{\vec k}}\hat{s}^{n{\vec m}}
\label{prediction}
\end{equation}
where $\hat{s}^\alpha=e^{-\widehat{w}^\alpha}$, $\widehat{w}^\alpha$ is the 
(complexified) volume of the holomorphic disc with boundary ending
on a particular one-cycle on the sL submanifold
and $q^a =e^{-t^a }$, $t^a $ being the closed string 
K\"ahler class. ${\vec m}$ are the winding numbers associated
with the holomorphic disc i.e., an element of $H_1$ of the sL
submanifold and ${\vec k}$ 
is an element of $H_2$ of the CY. The integer
$n$ relates to the multiple coverings of these.  
Further, this prediction involves the highly non-trivial integers
$d_{{\vec m},{\vec k}}$, that count the number of primitive disc
instantons in this homology class. These have the interpretation of
counting the degeneracy of domain wall D4-branes 
ending on the space-filling D6-brane\cite{ov}, which
wrap a 2-cycle in the CY (whose K\"ahler class is given by $\vec{k}$) and  
$\vec{m}$ denotes the wrapping number around the boundary.

Since a direct computation of the superpotential in the A-model
involves summing up
open-string instantons,  AV and AKV use the mirror map of \cite{hv}
to map this computation to one involving holomorphic Chern-Simons theory
in the mirror B-model.
For the case when the B-brane is of 
complex dimension one, and wraps a Riemann surface $C$ in the mirror CY, 
the dimensional 
reduction of the Chern-Simons theory on $C$ was computed in \cite{AV} 
in terms of the holomorphic three-form of the Chern-Simons theory 
restricted to the brane and two scalar fields (denoted by $u$ and
$v$ parametrising normal deformations of the B-brane in the mirror 
CY\footnote{These are $w^1$ and $w^2$, respectively,
in the notation of the previous subsection. For the convenience
of the reader, we shall use the convention of AKV in this subsection.}.
The final result for the superpotential 
is given by the the Abel-Jacobi map associated  with the one-form
$vdu$
\begin{equation}
W(u)=\int_{u^*}^u v(u) du
\end{equation}
where $u^*$ is the `location' of the half-brane. 
$v$ is fixed on the B-model side by the condition that $(u,v)$ be points
on the Riemann surface $C$.  However, from the A-model viewpoint, it
suffices to note that
\begin{equation}
v = \partial_{u} W(u)
\end{equation}
which is obtained from the Abel-Jacobi map. 

There are a couple of issues that one has to address before one can
translate these results in the B-model to the A-model side. First,
there is an $SL(2,\BZ)$ ambiguity in the parametrisation of the 
B-model variables given by
\begin{equation}
\pmatrix{u\cr v} \rightarrow \pmatrix{a & b \cr c & d} \pmatrix{u\cr v}
\quad.
\end{equation}
Thus, the superpotential given above is also dependent
on the choice of $u$ and $v$.
Second, the variables  which appear in the B-model side 
are not the same as the ones that appear in the A-model.  
The change of variables that relates the two is the
{\em open-string mirror map}. 

However, in the limit where quantum
corrections are suppressed in the A-model, i.e., where the volumes of the
holomorphic disc and the CY are large,  one can relate the two
sets of variables. However, the condition that Re($u$) be the volume of
the holomorphic disc and
$v\rightarrow 0$ in this limit, breaks the $SL(2,\BZ)$ down to the
subgroup given by
\begin{equation}
\pmatrix{u\cr v} \rightarrow \pmatrix{1 & n \cr 0 & 1} \pmatrix{u\cr v}
\quad.
\end{equation}
Thus, imposing 
the classical limit leaves us with an ambiguity parametrised
by the integer $n$. It turns out that for cases where the A-model is
dual to large $N$ Chern-Simons theory\cite{gv1,gv2,ov},
this integer reflects a framing ambiguity in the Chern-Simons theory.

The framing ambiguity has recently been studied in the context of the
Wilson loop variables of Chern-Simons theory in \cite{mv}, and evidence
has been found that the above mentioned duality is more appropriate in
the context of $U(N)$ Chern-Simons theories rather than the $SU(N)$ ones
originally considered in the proposal of \cite{ov}. Initial works on
Knot invariants were insensitive to this because the computations were
performed in the standard framing, where the $U(N)$ and $SU(N)$ Knot
invariants are the same. However, hints of this had already appeared in
\cite{lmv}. In \cite{mv}, explicit computations of certain $U(N)$ Wilson
loop operators have been carried out, and the results have been shown to
conform to the integrality predictions of \cite{ov}.

Aside from the ambiguity mentioned above, one has to obtain the
open-string mirror map in order to obtain the A-model results. It has
been argued in AKV that the appropriate variables are given by\cite{AKV}
\begin{eqnarray}
\hat{u} &=&  u + \Delta(\vec{q}) \\
\hat{v} &=& {{\partial W} \over{\partial \hat{u}}}
\end{eqnarray}
where $\hat{u}$ receives corrections only from closed-string 
instantons. $\hat{u}$ is computed by identifying it with the tension
of the domain wall which is mirror to the D4-brane in
the A-model. The variables $u$ and $v$ are conjugate fields of
the holomorphic Chern-Simons theory. The second equation corresponds
to the condition that $\hat{v}$ remains conjugate to $\hat{u}$ in
the quantum theory.

As we will see  in the next section, the B-model variables, i.e.,
the unhatted variables, naturally appear in the GLSM that we construct
to describe these half-branes.

\section{Linear Sigma Models for A-branes}
\subsection{The basic boundary condition}

The basic A-type boundary conditions that one considers 
in a GLSM with $n$ chiral superfields and $r$
$U(1)$ vector multiplets are
\begin{equation}
{{D^a}\over{e^2}} =r^a - \sum_i Q_i^a  |\phi_i|^2 = 0  \quad a
=1,\ldots,r
\end{equation}
The bosonic boundary conditions (in the matter sector)
are completed by the following $(n-r)$
conditions
\begin{equation}
\sum_i v^b_i\ {\rm Im} \log \phi_i   =0\quad b=1,\ldots,(n-r)
\end{equation}
where the {\em Lagrangian condition} implies that the $v$'s satisfy
$$
\sum_i v^b_i\ Q_i^a =0\quad. 
$$
This condition permits us to identify, in
the toric description, the $v^b_i$ with the integral generators of
the one-dimensional cones in the fan $\Delta$, the combinatoric
data  associated with the toric variety.

The boundary conditions on the vector multiplet are given by
the superfield condition
\begin{equation}
\Sigma_a' =0
\end{equation}
The easy way to understand this is to notice that in the topological
A-model, the field $\sigma$ is proportional to the K\"ahler form
$\omega$ whose restriction on a Lagrangian submanifold is zero.
The vanishing of the boundary terms in the ordinary variation
is obtained by the conditions
\begin{equation}
v_{01}^a =0
\end{equation}
The above set of boundary conditions form a {\em consistent} set of
boundary conditions in the GLSM which also have a proper NLSM limit
in the sense used in \cite{lgtwo,lsmone}.
Let us denote by $|{\cal B}\rangle_0$ the boundary state (in the GLSM)
that corresponds to these boundary conditions.

This reference state plays a role similar to the D6-brane in the context
of B-branes in the GLSM\cite{helices,lsmtwo}. 
It is not hard to see that it corresponds to
the D6-brane in the mirror Calabi-Yau manifold. It is also rigid
i.e., there are no moduli associated with this sL submanifold.

\subsection{The AV boundary conditions in the GLSM}

The boundary conditions that AV choose can be implemented in
the GLSM in a simple fashion\footnote{A similar construction was pursued
in \cite{hori} though it differs from ours in some details.}. 
In the matter sector, it corresponds
to simply replacing some of the $v$'s by boundary conditions of
the form\footnote{The angles dual these are given by
$\tilde{q}_\alpha^i {\rm Im} \log \phi_i$, where
$\tilde{q}_\alpha^i $ are the {\em dual basis vectors} (in $\BC^n$)
to the $q^{\alpha}_i$ i.e., they
satisfy: $\sum_i\tilde{q}_\alpha^i q^{\alpha\prime}_i
=\delta_\alpha^{\alpha\prime}$, $\sum_i\tilde{q}_\alpha^i Q^a_i=0$
and $\sum_i\tilde{q}_\alpha^i v_i^\beta=0$.}
$$
\sum_i q_i^\alpha |\phi_i|^2 = c^\alpha
$$
In the GLSM, we will realise these boundary conditions as {\em
low-energy conditions} using complex boundary supermultiplets
$U_\alpha$. The corresponding boundary state can be represented
by the path-integral 
\begin{equation}
|{\cal B}\rangle_{AV} = \int [DU]\ \exp(S_1+S_2+S_3)\ |{\cal B}\rangle_0
\end{equation}
where 
\begin{equation}
S_1 = \int dx^0 d^2\theta \left(\sum_i q_i^\alpha \bar{\Phi}'_i  
\Phi'_i   \right)
\left[ \left({{U_\alpha + \bar{U}_\alpha}\over2}\right) 
- \sum_j\tilde{q}_\alpha^j {\rm Im} \log \Phi_j' \right]
\end{equation}
This action is invariant under the gauge-invariances
\begin{eqnarray}
U_\alpha &\rightarrow& U_\alpha + L_\alpha \\
{\rm Im}\log \Phi'_i &\rightarrow& {\rm Im}\log \Phi'_i + 
q_i^\alpha \left({{L_\alpha +\bar{L}_\alpha}\over2}\right) \quad.
\end{eqnarray}
where $L_\alpha$ are chiral superfields. This gauge invariances
enable us to identify the real parts of the lowest components of the 
superfields $U_\alpha$ and with ``coordinates'' on an appropriate three-cycle. 
The action $S_1$ is also invariant under bulk gauge
transformations.  On using the fact that $D^a$ are generators of
the bulk gauge invariance, one obtains that 
$$
D^a |{\cal B}\rangle_{AV} =0\quad,
$$ 
given that $S_2$ and $S_3$ (given below)
are also gauge-invariant.

The kinetic energy for $\Xi_\alpha\equiv \bar{D} U_\alpha$
\begin{eqnarray}
S_2 &=& -{1\over{2e_b^2}}\int dx^0 d^2\theta\ \bar{\Xi}_\alpha \Xi_\alpha  
\nonumber \\
    &=& {1\over e_b^2}\int dx^0 \left(i\bar{\xi}_\alpha \partial_0 \xi_\alpha + 
       {1\over2} (D'_\alpha)^2 + 
{1\over2} (R_\alpha +\partial_0 u_\alpha)^2 \right)
\end{eqnarray}
where we have introduced the coupling constant $e_b$ on the boundary.
This has mass dimension half. 

We can now introduce the boundary analogue of the F.I. term:
\begin{equation}
S_3 = -{\rm Re}\left(w^\alpha \int dx^0 d\theta\ \Xi_\alpha\right)
=-\int dx^0 \left(c^\alpha D'_\alpha  - {\theta^\alpha\over{2\pi}}
(R_\alpha + \partial_0 u_\alpha)\right)
\end{equation}
where $w^\alpha=c^\alpha + i {\theta^\alpha\over{2\pi}}$ is a complex parameter. 

\subsection{Taking $e_b\rightarrow\infty$}

In the limit $e_b\rightarrow\infty$, the fields in the multiplet $\Xi_\alpha$
impose the constraints
\begin{eqnarray}
{D_\alpha'\over e_b^2}
=c^\alpha - \sum_i q_i^\alpha |\phi_i|^2 =0 \nonumber \\
\sum_i q_i^\alpha \left(\overline{\phi}_i \psi_{+i} + \eta \phi_i
\overline{\psi}_{-i} \right)=0  \\
\sum_i q_i^\alpha \left(\overline{\phi}_i \psi_{-i} + \eta \phi_i
\overline{\psi}_{+i}\right) =0 \nonumber  \\
\eta \sum_i q_i^\alpha (\bar{\phi}_i F_i + \phi_i F^*_i)
+i\sum_i q_i^\alpha 
(\bar{\phi}_i D_1 \phi_i - \phi_i D_1 \bar{\phi}_i) =0\quad. \nonumber 
\end{eqnarray}
The last three equations are
the supersymmetric variation of the first equation.

The $\theta^\alpha$ term is a topological term as one expects.
Thus, we can see that the real part of $R$ behaves like the bulk $D$ term 
while the imaginary part should behave like $v_{01}$. In order for this
correspondence to work, we should require that (after Wick rotation
to Euclidean space and $x^0$ becomes a circle variable which we
will call $x^2$ with periodicity $2\pi$)
$$
\int {{dx^2}\over{2\pi}} (R_\alpha + \partial_2 u_\alpha) 
= -m_\alpha
$$
where $m_\alpha$ are some integers.

\section{The topological version of the GLSM}

In order to discuss the topological aspects of the A-model, we need to
Wick rotate: $x^0\rightarrow-ix^2$ (Hence one has
$\partial_0\rightarrow i\partial_2$, $R\rightarrow iR$
and $v_{01}\rightarrow -i v_{12}$). Let us choose to twist the
model so that $\bar{Q}$ is the BRST operator. For the closed
string case, the localisation equations are\cite{wittenphases} 
$(a=1,\cdots,h_{1,1})$.
\begin{eqnarray}
(D_1 + i D_2) \phi =0 \nonumber \\
D^a + v_{12}^a =0 \\
F_i =0 \nonumber \\
\sigma^a =0 \nonumber
\end{eqnarray} 
Let $n_a$ be the instanton numbers defined by
\begin{equation}
n_a \equiv -{1\over{2\pi}} \int_D d^2x\ v_{12}^a
\end{equation}
The action is then equal to 
\begin{equation}
S_{\rm bulk} = -2\pi i t^a n_a
\end{equation}
where $t^a = {\theta^a\over{2\pi}} + i r^a$.

In the open-string sector, one obtains
\begin{eqnarray}
D_\alpha' + (R_\alpha + \partial_2 u_\alpha) =0 \nonumber \\
\xi_\alpha =0
\end{eqnarray}
in addition to the boundary condition
\begin{equation}
\sum_i v^\beta_i {\rm Im} \log \phi_i =0\quad.
\end{equation}
The winding numbers are defined by
\begin{equation}
m_\alpha = -{1\over{2\pi}}\int_{\partial D} (R_\alpha + \partial_2
u_\alpha)
\end{equation}
The contribution of the boundary terms to the action is
\begin{equation}
S_{\rm bdry}  = -2\pi i s^\alpha m_\alpha
\end{equation}
Thus, the total action in the topological theory is given by
$$
S = S_{\rm bulk} + S_{\rm bdry} = -2\pi i t^a n_a -2\pi i s^\alpha
m_\alpha\quad.
$$
\subsection{Observables in the topological theory}

The gauge-invariant BRST closed
observables in the closed string sector are
$\sigma^a$. These have ghost number two. In the open-string sector,
the observables in the NLSM limit should be in one-to-one correspondence
with closed forms on the Lagrangian submanifold. In the AV construction,
we assume that a subset of closed one-forms  is given by $du_\alpha$,
where we consider $u_\alpha$ to be (compact) coordinates (angles)
on the Lagrangian submanifold. Thus, in the topological model, the
corresponding observables are $\xi_\alpha$ (which are naively BRST
exact). Thus one has
\begin{center}
\begin{tabular}{|c|c|c|}\hline
 & Observable & Ghost No. \\ \hline
Bulk & $\sigma^a$ & $2$ \\ \hline
Boundary & $\xi_\alpha$ & $1$ \\ \hline
\end{tabular}
\end{center}

\section{The open-string instanton moduli space}

	The open-string instanton moduli space can be defined to be
the moduli space of the localisation equations that we derived in
the previous section. We assume that the worldsheet is a disc.
The relevant equations are:
\begin{eqnarray}
(D_1 + i D_2) \phi_i =0  \\
D^a + v_{12}^a =0 \label{bulk}\\
\left. D_\alpha' + (R_\alpha + \partial_2 u_\alpha)\right|_{\partial D}
\label{bounda} =0 \\
 \sum_i v_i^\beta {\rm Im} \log \phi_i  |_{\partial D}
\label{boundb} =0
\end{eqnarray}
The first two equations are bulk equations while the last two equations
are on the boundary of the disc.  Thus, we are interested in the
moduli space of these equations modulo bulk as well as boundary (real)
gauge invariances. Further, the instanton moduli spaces are fixed by
the instanton numbers $n_a$ as well as the winding numbers $m_\alpha$.

The first equation is invariant under {\em complex} gauge
transformations. As in the bulk case, we shall consider the last two
equations as gauge-fixing of the imaginary part of the gauge
transformation. Thus, we shall be interested in the moduli space of
the equations
\begin{equation}
(D_1 + i D_2) \phi_i =0  
\end{equation}
modulo {\em complex} gauge invariances. Under suitable circumstances
i.e., when one is dealing with a {\em stable} situation,
this moduli space is identical to
the moduli space  of the full set of equations including
(\ref{bulk}) and (\ref{bounda}) modulo {\em real} gauge invariance.
In a suitable gauge where $v_{\bar{z}}^a=0$, this
equation states that $\phi_i$ is a holomorphic function on the disc.
In the closed string case, this implies that $\phi_i$ are
sections of $O(n_i\equiv \sum_aQ_i^an_a)$ in the sector with instanton
number $n_a$.

\subsection{Stability}

In order to obtain the conditions under which stability is obtained
(see for instance \cite{bradlow}),
we shall first consider the bulk equation (\ref{bulk}). Let us choose
the worldsheet be a disc $D$. Integrating (\ref{bulk}) over the disc we
obtain
\begin{equation}
2\pi n^a +e^2 \int_D Q_i^a |\phi_i|^2 = e^2 r^a {\rm vol}(D)
\end{equation}
Let us consider the case when $e^2r^a\gg 0$ (for some fixed $a$)
and all fields $\phi_i$
whose $U(1)$ charge is negative are set to {\em zero}. One then
obtains the following inequality
\begin{equation}
{\rm vol}(D) \geq {{2\pi n^a}\over{e^2 r^a}} 
\end{equation}
It is not hard to see that for fixed $n^a\geq0$, this condition can always
be satisfied for large enough $e^2 r^a$. Thus stability is guaranteed
in this case.
A similar argument occurs for the boundary equation (\ref{bounda})
leads to the condition
\begin{equation}
{\rm vol}(\partial D) \geq {{2\pi m_\alpha}\over{e_b^2 c^\alpha}} 
\end{equation}
where stability is again guaranteed for large enough $e_b^2 c^\alpha$
for $m_\alpha\geq0$. The conditions involving the phases (eqn.
(\ref{boundb})) can always be imposed and hence do not play a role
in considerations of stability.

Let us now consider the situation where the disc is large i.e., can
be approximated by the complex plane. Then, one has the following
interpretation for the bulk instanton numbers and the boundary
winding numbers. Consider $n_i\equiv Q_i^a n^a + q_i^\alpha
m_\alpha$. Then, one has (as solutions of the localisation equations)
\begin{enumerate}
\item $\phi_i=0$ whenever $n_i<0$.
\item $\phi_i= \sum_{j=0}^{n_i} \phi_{i,j}\ z^j$ otherwise
\end{enumerate}
where $z$ is a (complex) coordinate on the disc. The winding numbers
(instanton numbers) fix the leading power of $z$ (number of zeros on
the disc). Let us denote this space by $Y_{n^a,m_\alpha}$.

Thus, the moduli space is given by the space of $\phi_{i,j}$ for all
$n_i>0$ modulo gauge invariances. This is fixed by
\begin{enumerate}
\item The bulk gauge invariance is fixed by a D-term of the form
$$
\sum_i Q^a_i \left(\sum_{j=0}^{n_i} |\phi_{i,j}|^2\right)  = r^a
$$
and the associated $U(1)$ gauge invariance.
\item The boundary condition on the disc, eqn. (\ref{bounda}),
becomes a constraint of the form
$$
\sum_i q^\alpha_i \left(\sum_{j=0}^{n_i} |\phi_{i,j}|^2\right)  = c^\alpha
$$
\item The boundary conditions in (\ref{boundb}) can be considered to
arise from the gauge fixing of the $U(1)$ gauge invariance 
$$
\phi_{i,j} \rightarrow e^{iv_i^\beta \theta} \phi_{i,j}\quad.
$$
\end{enumerate}
In this case, the moduli space is the variety given by 
\begin{equation}
{\cal M}_{n_a,m_\alpha} = (Y_{n_a,m_\alpha} - F_{n_a,m_\alpha})/T
\end{equation}
where $T$ acts as 
\begin{eqnarray}
\phi_{i,j} &\rightarrow& \lambda^{Q_i^a} \
\phi_{i,j}\quad{\rm for}~ \lambda\in \BC^* \quad, \\
\phi_{i,j} &\rightarrow& \lambda^{q_i^\alpha}\
\phi_{i,j}\quad{\rm for}~\lambda\in \BR^+ \\
\phi_{i,j} &\rightarrow& e^{iv_i^\beta \theta}\ \phi_{i,j}
\quad{\rm for~real}~\theta \quad.
\end{eqnarray}
$F_{n_a,m_\alpha}$ is the set of
points in $Y_{n_a,m_\alpha}$ for which the D-term constraints
given above are not satisfied. Note that in the zero instanton sector
(i.e., $n_a=0,m_\alpha=0$), one obtains the sL submanifold as expected.

\subsection{The boundary toric variety}

Suppose, we treat the D-term-like  boundary conditions
which are specified by $q_i^\alpha$ on par with the bulk
D-terms specified by $Q_i^a$. The special Lagrangian condition
$\sum_i q_i^\alpha=0$ now becomes the Calabi-Yau condition on
a toric-variety which we shall call the {\em boundary toric variety}.
The boundary moduli now become bulk moduli on the toric variety.
One can see that the spaces $Y_{n_a,m_\alpha}$ and ${\cal M}_{n_a,m_\alpha}$
obtained in the previous subsection play a similar role for the boundary 
toric variety. A well known method of computing mirror periods of a 
Calabi-Yau hypersurface is by the Picard-Fuchs system of differential 
equations. By treating the bulk and the boundary D-term constraints on 
the same footing, one might hope to compute the period integrals of the 
boundary toric variety in a similar fashion. This should, in principle, 
directly give the necessary variables for the computation of the 
worldvolume superpotential for A-branes. In the following sections, we 
will proceed with such a computation, and show that the boundary toric
variety indeed reproduces the structure of the A-model amplitude 
predicted in \cite{ov}.

\section{Summing up open-string instantons}

\subsection{The GKZ equation for the bulk GLSM}

As discussed earlier, the toric data is seen in the GLSM 
construction through the $U(1)$ charges, $Q_i^a$ ($a=1,\ldots,r$), 
of the chiral superfields and $v_i^b$ ($b=1,\ldots,(n-r)$) which 
are the integral generators of
one-dimensional cones in the fan $\Delta$ of the toric variety.
Consider the differential operators constructed from this data
\begin{eqnarray}
{\cal L}_{Q^a} &=& \prod_{Q^a_i>0}\left(\frac{\partial}{\partial
a_i}\right)^{Q^a_i}
-\prod_{Q^a_i<0}\left(\frac{\partial}{\partial a_i}\right)^{-Q^a_i}
\label{gkza} \\
Z_b&=&\sum_{i=0}^{n}v_i^b a_i\frac{\partial}{\partial a_i}-\hat{\beta}_b
\end{eqnarray}
with $\hat{\beta}=(-1,0,\ldots,0)$ and $a_i$ are coordinates on $\BC^n$.
The two operators given above define a consistent system of 
differential equations,  due to Gelfand, Kapranov and Zelevinsky, 
the GKZ hypergeometric system with exponent $\hat{\beta}$ 
\begin{equation}
{\cal L}_{Q^a}\ \tilde{\pi}(a)=0\quad,\quad 
Z_b\ \tilde{\pi}(a)=0
\end{equation}

It is known that the periods (associated with three-cycles) of the
mirror Calabi-Yau manifold are solutions of a
Picard-Fuchs differential equation which naturally arises in the special
geometry associated with vector multiplets with $N=2$ supersymmetry
in four dimensions. It turns out that for the CY manifolds which arise
as hypersurfaces in toric varieties, the GKZ equations described
above are closely related to these Picard-Fuchs equations. 
In the cases that we will be interested in, the above system of
equations provide the Picard-Fuchs differential equations. In general
however, one may need to supplement the above system by additional
differential operators to convert the GKZ into the relevant
Picard-Fuchs equation.

The second set of differential equations, i.e., those given by the
$Z_b$, are  solved for by a change of variables $a_i\rightarrow z_a$
given by
$$
z_a = \prod_i (a_i)^{Q^a_i} \quad,
$$
and the change $\tilde{\pi}(a)={1\over a_0} \pi(z)$. Further, it is
useful
to rewrite the differential operators ${\cal L}_{Q^a}$ in terms of
logarithmic derivatives $\Theta_a = z_a {\partial\over{\partial z_a}}$.

Let us assume that $z_a\rightarrow 0$ is the large-volume limit.
There are two classes of solutions (periods) of the GKZ equation which
are
of interest (see \cite{yauetal} for the case of compact CY manifolds
and \cite{chiangetal} for the non-compact case [local mirror symmetry]).
\begin{enumerate}
\item Those with logarithmic behaviour as $z\rightarrow 0$. These are
important in providing the change of variables which 
form the {\em mirror map}.
This also relates the
the GLSM variables with the NLSM (large-volume) variables. This is
clearly seen in the work of Morrison and Plesser\cite{morple}.
\item Those that behave as $(\log z)^2$ as $z\rightarrow 0$. Using
special geometry, these are related to the first derivatives of
the prepotential. This prepotential encodes the corrections due
to closed-string instantons to the classical one obtained from
the triple intersection numbers of two-cycles on the CY manifold.
\end{enumerate}

\subsection{The GKZ system for the LSM with boundary}

We will now show that we can sensibly fit the case of LSM with boundary
into this framework and write down the corresponding GKZ equations. We
will explicitly show this for all the cases that have been considered by
AV and AKV. Loosely speaking, the AV boundary conditions correspond 
to replacing some of the $v$'s with $q$'s. This suggests that the GKZ
obtained by the set of charge vectors $(Q_i^a,q_i^\alpha)$ might be
relevant for obtaining the open-string instanton effects just as
the bulk GKZ for the closed-string case. Consider the following
operators
\begin{eqnarray}
{\cal L}_{Q^a} &=& \prod_{Q^a_i>0}\left(\frac{\partial}{\partial
a_i}\right)^{Q^a_i}
-\prod_{Q^a_i<0}\left(\frac{\partial}{\partial a_i}\right)^{-Q^a_i}
 \\
{\cal L}_{q^\alpha} &=& \prod_{q^\alpha_i>0}\left(\frac{\partial}{\partial
a_i}\right)^{q^\alpha_i}
-\prod_{q^\alpha_i<0}\left(\frac{\partial}{\partial a_i}\right)^{-q^\alpha_i}
\\
Z_\beta&=&\sum_{i=0}^{n}v_i^\beta a_i\frac{\partial}{\partial
a_i}-\hat{\beta}_\beta
\end{eqnarray}
This is the GKZ which
is naturally associated with the boundary toric variety. 
As before, we will
solve the differential equations given by $Z_\beta$ by a change of
variables
$$
z_a = \prod_i (a_i)^{Q^a_i} \quad, \quad
s_\alpha = \prod_i (a_i)^{q^\alpha_i} \quad, 
$$
where $s_\alpha$ are related to boundary moduli just as the $z_a$ are
related to bulk moduli.
It is useful
to separate the remaining differential operators into those
that arise from the $Q_i^a$ (we will call these ${\cal L}_{\rm bulk}$)
and those that arise from the $q_i^\alpha$ (we will call these
${\cal L}_{\rm bdry}$). Note that ${\cal L}_{\rm bulk}$ is not quite
the differential operator which appears in the closed-string GKZ --
it also involves the boundary variables.

\subsection{Identifying the solutions of interest}

We will restrict our attention to the cases of the half-branes
with topology $\BC\times S^1$ and let $\hat{s}=\exp(-\widehat{w})$
be the modulus associated with the $S^1$. The limit where quantum
corrections are small is given by $\vec{z},s\rightarrow 0$.

By construction, the bulk periods are solutions of the differential
equations involving ${\cal L}_{\rm bulk}$. In particular, the solutions
that provide the {\em closed-string mirror map} are solutions with
asymptotic behaviour $\log z$ as $z,s\rightarrow 0$. The AKV ansatz
plays a similar role in obtaining the {\em open-string mirror map}
i.e., we look for solutions of ${\cal L}_{\rm bulk}$ with the
property 
$$
\widehat{w} =  \log s + \Delta(\vec{z}) 
$$
It is not hard to see that the above ansatz is not compatible with
${\cal L}_{\rm bdry}$ since $\Delta$ is not a function of $s$. 

The solution associated with the superpotential is obtained by
an ansatz of the form
$$
\partial_{\widehat{w}} W(\hat{w}) = 
\sum_{\vec{k},m} a(\vec{k},m)\ z^{\vec k} s^m
$$
where we require that the zero instanton contribution vanish i.e.,
$a(0,0)=0$
and the only non-vanishing $a(\vec{k},m)$ are as given by the
considerations of section 5. To be more precise, we set $a(\vec{k},m)=0$
whenever the corresponding instanton moduli space does not exist.
Unlike the solution associated with the open-string mirror map
we require that this solves the full boundary GKZ differential equation.

\section{Examples}

\subsection{${\cal O}(-3)$ on $\BP^2$}

\subsubsection{Bulk Periods}

The GKZ equation is given by
\begin{equation}
\left[\Theta_z^3 + z (3 \Theta_z + 2) (3\Theta_z +1) 3\Theta_z \right]\ \pi =0
\end{equation} 
The basic solutions which we consider are fixed by the behaviour near $z=0$:
(i) the constant solution; (ii) the $\log z$ solution and (iii) the $(\log z)^2$
solution. Using the Frobenius method, one obtains
\begin{eqnarray}
\pi_0 &=&1 \\
\pi_1 &=& \log z + \sum_{n=1}^\infty (-)^n{3n!\over{(n!)^3}}  {z^n \over n} \\
\pi_2 &=& {1\over2}\pi_1^2 + {\cal C}^2 + \sum_{n=1}^\infty (-)^n {3n!\over{(n!)^3}}
\left(-{1\over n} + \sum_{j=n+1}^{3n} {1\over j} \right) {z^n \over n}
\end{eqnarray}
where ${\cal C}=\sum_{n=1}^\infty (-)^n{3n!\over{(n!)^3}}  {z^n \over n}$.

The mirror map relates the large-volume coordinate $t$ to $\pi_1$. Thus, we
obtain
\begin{equation}
t = -\pi_1/\pi_0 = 
-\log z - \sum_{n=1}^\infty (-)^n{3n!\over{(n!)^3}}  {z^n \over n}
\end{equation}
which has the right behaviour $z=e^{-t}$ as $z\rightarrow0$ and the terms
in the summation are instanton corrections. The inverse relation is of the form
\begin{equation}
z = q + 6 q^2 + 9 q^3 + 56 q^4 -300 q^5 + 3942 q^6 -48412 q^7 + \cdots
\label{closedflat}
\end{equation}
where $q\equiv e^{-t}$

\subsubsection{Phase I}

The GKZ equation for the boundary toric variety is given by
\begin{eqnarray}
\left[\Theta_z^2 (\Theta_z + \Theta_s) + z (3 \Theta_z + \Theta_s + 2) 
(3\Theta_z + \Theta_s +1) (3\Theta_z + \Theta_s) \right]\ \pi =0 \\
\left[(\Theta_z + \Theta_s) +s (3\Theta_z + \Theta_s)\right]\ \pi =0
\end{eqnarray} 
The first equation represents ${\cal L}_{\rm bulk}$ and the
second represents ${\cal L}_{\rm bdry}$ for this example. Since $c^2=0$
in this phase, we have ``switched-off'' the coordinate associated
with $c^2$.

The boundary flat coordinate $\hat{u}$ satisfies the first of the two equations given
above. For $z,s\rightarrow 0$, one requires that $\hat{u} = - \log s$. It can
be argued that the general solution should take the form
$$
\hat{u} = -\log s + \Delta(z)
$$
i.e., the boundary flat coordinate differs from the GLSM coordinate only by
closed moduli. One can then easily prove that
$$
\hat{u} = -\log s - {1\over3} (t- \hat{t} ) + i\pi
$$
is a solution of the first of the two equations 
(i.e., ${\cal L}_{\rm bulk}$) that make up the boundary GKZ equations.

Defining $\hat{s}=e^{-\hat{u}}$, one obtains the following
\begin{equation}
s = -\hat{s}  (1 + 2 q - q^2 + 20 q^3 - 177 q^4 + 1980 q^5 - 24023 q^6 + 
    97684 q^7+\cdots) 
\label{openflat}
\end{equation}
This equation along with the earlier equation for $z$, gives us the required
change of variables from $z,s$ to the flat coordinates $\hat{s}$ and $q$.

We are now in a position to compute the boundary superpotential. We 
require that it be a solution $\hat{v}$
of both equations of the boundary GKZ. It should satisfy the following 
conditions: (a) $\hat{v}$ should vanish and be regular at $z=s=0$.
(b) By consider the instantons in the GLSM, one obtains the vector
$kQ + m q^1=(-3k-m,k+m,k,k)$. In phase (i), we need both $\phi_1$ and $\phi_3$ to be 
non-vanishing. This is achieved if $k\geq0$ and $k+m\geq0$. This suggests the
following ansatz for this solution:
$$
\hat{v} = \sum_{k=0}^\infty \sum_{m=-k}^\infty a(k,m) z^k s^m
$$
It may seem that this solution cannot be regular at $z=s=0$. 
But, since we require
that $c^1 < r$, this requires the limit be take such that $z/s\rightarrow 0$.
Thus $z^k s^m$ vanishes for $z=s=0$ as long as $k+m\geq0$.

The solution is given by 
\begin{equation}
\hat{v} = \sum_{k=0}^\infty \sum_{m=-k}^{\infty} (-1)^{k+m} 
{\Gamma(3k+m) \over{\Gamma(k+m+1)\Gamma(k+1)^2}} z^k s^m
\end{equation}

Putting in the expressions for the flat coordinates, by making the 
substitutions as in (\ref{closedflat}) and (\ref{openflat}), and comparing
with the expression 
\begin{equation}
{\hat v}=-\sum_{k,m}md_{k,m}\log\left(1-q^k{\hat s}^m\right)=
\sum_{k,m,n}\frac{m}{n}d_{k,m}\left(q^k{\hat s}^m\right)^n\quad.
\end{equation}
Table 1 lists the degeneracies $d_{k,m}$ of disc instantons.
\begin{table}
$$
\begin{array}{|c|rrrrrr|}\hline
m\backslash k &~0 &1 &2 &3 &4 &5 \\ \hline
-5& 0& 0& 0& 0& 0& 5
\\ -4& 0& 0& 0& 0& -2& 28
\\ -3& 0& 0& 0& 1& -10& 102
\\ -2& 0& 0& -1& 4& -32& 326
\\ -1& 0& 1& -2& 12& -104& 1085
\\ 1& 1& -1& 5& -40& 399& -4524
\\ 2& 0& -1& 7& -61& 648& -7661
\\ 3& 0& -1& 9& -93& 1070& -13257
\\ 4& 0& -1& 12& -140& 1750& -22955
\\ 5& 0& -1& 15& -206& 2821& -39315
\\ 6& 0& -1& 19& -296& 4450& -66213
\\ 7& 0& -1& 23& -416& 6868& -109367
\\ \hline
\end{array}
$$
\caption{Degeneracies for phase I of ${\cal O}(-3)$ on $\BP^2$}
\end{table}

\subsubsection{Phase III}
The charges relevant for this phase are:
$Q=(-3,1,1,1)$ and $q=-q^2=(1,0,-1,0)$.
The relevant GKZ equations are 
\begin{eqnarray}
&~&\left[\Theta_{z}^2\left(\Theta_{z}-\Theta_{r}\right)
+z\left(3\Theta_{z}-\Theta_{r}+2\right)
\left(3\Theta_{z}-\Theta_{r}+1\right)
\left(3\Theta_{z}-\Theta_{r}\right)\right]\ \pi=0\nonumber\\
&~&\left[
\left(3\Theta_{z} -\Theta_{r}\right)+
r \left(\Theta_{z}-\Theta_{r}\right)
\right] \ \pi=0
\label{gkz3}
\end{eqnarray}
where the boundary modulus is represented by $r$ in this phase.
In this case, as before, the boundary flat coordinate ${\hat v}$ has a 
general solution
\begin{equation}
{\hat v}=\log\ r +\frac{1}{3}(t-{\hat t}) +i\pi
\end{equation}
and defining $\hat{r}=e^{{\hat v}}$, we have the expansion
\begin{equation}
r=-{\hat r}\left[1 - 2q + 5q^2 - 32q^3 + 286q^4 -3038q^5 +35870q^6+
\cdots\right] 
\end{equation}
The allowed instanton numbers in the GLSM are given by looking at
$kQ+mq=(m-3k,k,k-m,k)$. Since $\phi_0\neq0$ in this phase, we need
$m-3k\geq0$ in addition to $k\geq0$ which suggests an ansatz of
the form (for the superpotential)
\begin{equation}
{\hat u'}=\sum_{k=0}^{\infty} \sum_{m=3k}^\infty a(k,m)\ z^k\ r^m
\end{equation}
As can be easily seen, the GKZ equations (\ref{gkz3}) are solved 
with the choice of $a(k,m)$ as 
\begin{equation}
a(k,m)=(-)^{k+m}\frac{\Gamma(m-k)}{\Gamma(1-3k+m)\Gamma(k+1)^2}
\end{equation} 
The degeneracies that are obtained from these can be seen to be
consistent with those obtained by AKV (see Table 2).
\begin{table}
$$
\begin{array}{|c|rrrrrr|}\hline
m\backslash k &~0 &1 &2 &3 &4 &5  \\ \hline
\vdots & \vdots &&&\vdots&&\vdots     \\
   3& 0& -1& 3& -18& 153& -1560  
\\ 4& 0& -1& 4& -20& 160& -1595 
\\ 5& 0& -1& 5& -26& 196& -1875 
\\ 6& 0& -1& 7& -36& 260& -2403 
\\ 7& 0& -1& 9& -52& 365& -3254 
\\ 8& 0& -1& 12& -76& 528& -4578 
\\ 9& 0& -1& 15& -111& 784& -6627 
\\ 10& 0& -1& 19& -161& 1176& -9800 
\\ 11& 0& -1& 23& -230& 1777& -14720 
\\ 12& 0& -1& 28& -323& 2678& -22347 
\\ 13& 0& -1& 33& -446& 4018& -34141 
\\ 14& 0& -1& 39& -605& 5968& -52289 
\\
\vdots & \vdots &&&\vdots&&\vdots     \\ \hline
\end{array}
$$
\caption{Degeneracies for phase III of ${\cal O}(-3)$ on $\BP^2$}
\end{table}

\subsection{${\cal O}(-1)+{\cal O}(-1)$ on $\BP^1$}

The boundary toric variety is characterised by the charges
$Q=\left(1,1,-1,-1\right)$, and 
$q=\left(0,0,1,-1\right)$. Let us consider the phase where 
$c^1=0$ and $c^2>0$(phase II of AV). The relevant GKZ equations can 
easily be 
seen to be
\begin{eqnarray}
&~&\left[\Theta_{z}^2-z\left(\Theta_{z}+\Theta_{s}\right)
\left(\Theta_{z}-\Theta_{s}\right)\right]\pi=0\nonumber\\
&~&\left[\left(\Theta_{s}-\Theta_{z}\right)+s\left(\Theta_{s}
+\Theta_{z}\right)\right]\pi=0
\end{eqnarray}

Where, as before, $z=e^{-t}$ and we define $s=e^{v}$
In this example, there are no corrections to the open string moduli 
coming from closed string ones, and one can simply write 
\begin{equation}
{\hat v}=v + i\pi~,~~~~~~{\hat u}=u + i\pi
\end{equation}
so that ${\hat s}=-s$. In this case, the allowed instanton numbers 
are given by the vector 
$kQ+mq=\left(k,k,-k+m,-k-m\right)$, and since $\phi_3\neq 0$, we would 
require $m>k$, which suggests an ansatz of the form
\begin{equation}
\hat{u} = \sum_{k=0}^\infty \sum_{m=k}^\infty a(k,m)z^ks^m
\end{equation}
The appropriate $a(k,m)$ are easy to find, and, defining 
$z=e^{-t}$, the solution of 
the GKZ equations can be seen to be 
\begin{equation}
{\hat u}=\sum_{k=0}^{\infty}\sum_{m=k}^{\infty}
(-1)^{k+m}\frac{\Gamma(k+m)}{\Gamma(m-k+1)\Gamma(k+1)^2}z^k s^m 
\end{equation}

We now reproduce some of the degeneracies $d_{k,m}$ for this
example in  Table 3. The numbers agree with those in
\cite{AV} up to an overall sign.
\begin{table}
$$
\begin{array}{|c|rrrrrr|}\hline
m\backslash k &~0 &1 &2 &3 &4 &5  \\ \hline
\vdots & \vdots &&&\vdots&&\vdots     \\
11& 0& 1& -30& 390& -2730& 11466
\\ 12& 0& 1& -36& 556& -4690& 24024 
\\ 13& 0& 1& -42& 770& -7700& 47124
\\ 14& 0& 1& -49& 1040& -12152& 87516
\\ 15& 0& 1& -56& 1375& -18564& 155195
\\ 16& 0& 1& -64& 1785& -27552& 264537
\\ 17& 0& 1& -72& 2280& -39900& 435708
\\ 18& 0& 1& -81& 2871& -56520& 696388
\\ 19& 0& 1& -90& 3570& -78540& 1083852
\\ 20& 0& 1& -100& 4389& -107255& 1647455
\\ 21& 0& 1& -110& 5341& -144210& 2451570
\\ 22& 0& 1& -121& 6440& -191180& 3579030
\\ 
\vdots & \vdots &&&\vdots&&\vdots     \\ \hline
\end{array}
$$
\caption{Degeneracies for phase II of ${\cal O}(-1)+{\cal O}(-1)$ on
$\BP^1$}
\end{table}
\subsection{Degenerate Limit of $\BP^1\times\BP^1$}

In this subsection, we will calculate the superpotential for
another example, that of a non-compact Calabi-Yau, containing 
$\BP^1\times\BP^1$. We will consider the theory with one of the
closed string moduli goes to infinity, and will consider the phase
where the relevant charges that defines the other closed string 
modulus and Lagrangian submanifold are $Q=(-2,1,1,0,0), 
q=(-1,0,0,1,0)$. The GKZ equations for the boundary toric variety
defined by these charges are
\begin{eqnarray}
&~&\left[\Theta_{z}^2
-z\left(2\Theta_{z}+\Theta_{s}+1\right)
\left(2\Theta_{z}+\Theta_{s}\right)\right]\pi=0\nonumber\\
&~&\left[\Theta_{s}+s\left(2\Theta_{z}
+\Theta_{s}\right)\right]\pi=0
\end{eqnarray}
where we have defined $z=e^{-t}$ and $s=e^{v}$. The mirror map 
dictates that the boundary flat coordinate in this case is given by
\begin{equation}
{\hat s}=-(1+q)s
\end{equation}
where ${\hat s}=e^{{\hat v}}$, and similarly for the other open string
modulus. $q=e^{-{\hat t}}$ is defined through 
\begin{equation}
z=\frac{q}{(1+q)^2}
\end{equation}
  
The allowed instanton numbers are determined from the vector 
$kQ+mq=(-2k-m,k,k,m)$ and since $\phi_1,\phi_2,\phi_4$ are non-vanishing, 
we require that $k,m\geq 0$. Writing an ansatz for the solution to the
GKZ equations, 
\begin{equation}
{\hat u}=\sum_{k=0,m=0}^{\infty}a(k,m)z^ks^{m}
\label{udeg}
\end{equation}
The coefficients are easily determined to be
\begin{equation}
a(k,m)=(-)^{m}\frac{\Gamma(2k+m)}{\Gamma(m+1)\Gamma(k+1)^2}
\end{equation}
Substituting in (\ref{udeg}), with the summation excluding $(0,0)$, 
and, substituting for $z$ and $s$ in terms of $q$ and ${\hat s}$, we 
recover the integers $d_{k,m}$ which are consistent with the results
of AKV (see Table 4).
\begin{table}
$$
\begin{array}{|c|rrrrrr|}\hline
m\backslash k &~0 &1 &2 &3 &4 &5  \\ \hline
\vdots & \vdots &&&\vdots&&\vdots     \\
   11& 0& 1& 36& 676& 8281& 74529 
\\ 12& 0& 1& 42& 920& 12936& 132496 
\\ 13& 0& 1& 49& 1225& 19600& 226576 
\\ 14& 0& 1& 56& 1600& 28896& 374544 
\\ 15& 0& 1& 64& 2055& 41616& 600935 
\\ 16& 0& 1& 72& 2601& 58680& 938961 
\\ 17& 0& 1& 81& 3249& 81225& 1432809 
\\ 18& 0& 1& 90& 4011& 110550& 2140369 
\\ 19& 0& 1& 100& 4900& 148225& 3136441 
\\ 20& 0& 1& 110& 5929& 196020& 4516475 
\\ 21& 0& 1& 121& 7112& 256036& 6400900 
\\ 22& 0& 1& 132& 8464& 330616& 8940100 
\\
\vdots & \vdots &&&\vdots&&\vdots     \\ \hline
\end{array}
$$
\caption{Degeneracies for the degeneration of $\BP^1\times\BP^1$}
\end{table}

\section{Conclusion}

In this paper, we have constructed a GLSM with boundary
which corresponds to D-branes associated
with the  class of special Lagrangian cycles proposed by Aganagic and
Vafa.  The considerations of the topological field theory associated
with the GLSM naturally leads to the an associated  boundary toric
variety. We have seen that the  GKZ associated with the boundary toric
variety provides us with the {\em open-string mirror map} and enables
us to verify the integrality conjectures for the degeneracies
associated with open-string
instantons. We also obtain natural models for the open-string
instanton moduli spaces. 
We have however not directly computed the degeneracies in the
topological model -- this involves a generalisation to the GLSM with
boundary, of the methods
used by Morrison and Plesser for the closed-string case\cite{morple}. 
The methods used in this paper clearly have a generalisation to cases
involving compact CY manifolds i.e.,
those that include a bulk superpotential in the GLSM. However, there
might be cases where the moduli are related to non-toric deformations
for which special techniques are necessary.

The framing ambiguity is quite similar to what may happen in
the case of a conifold
singularity, in say the quintic. In that case, one has an $S^3$ which
shrinks to zero size. Its dual three-cycle is a $T^3$. The monodromy
action around the conifold singularity generates the same subgroup of
$SL(2,Z)$ that we saw in the framing ambiguity. The volume of the $T^3$
thus suffers from a similar ambiguity in the sense that one can always
add an integer times the $S^3$ volume (which clearly vanishes at the
conifold singularity) to this volume. However, in the
case of the quintic, this ambiguity can be fixed by global
considerations. In a similar fashion, we expect the framing ambiguity to
be associated with a monodromy of the GKZ solutions and that in the
compact case, this ambiguity should be lifted by global
considerations.

A `derivation' of the GKZ equations that we use can be obtained 
in a fashion similar to the closed-string case. 
By extending the Hori-Vafa \cite{hv} mirror map, one can represent the periods
on the B-model side as a functional integral over some dual fields with
appropriate superpotentials corresponding to all the D-terms, bulk
and boundary, that appear on the A-model side. Using this
representation, it is possible to derive the appropriate GKZ in a manner
similar to that of Hori and Vafa for the closed-string periods. 
From a somewhat different view point, one notices that 
the periods can be represented 
as an integral of the holomorphic
three-form over the three-cycle and then one may obtain the differential
equation satisfied by the periods. 
The GKZ equations that we use can be derived by
replacing the integral over the holomorphic  three-form by the
holomorphic Chern-Simons action. 

The AV boundary conditions that we have implemented in the
GLSM correspond to  two-branes on the B-model
side given by intersections of hyperplanes. It is of interest to extend
our methods to construct the GLSM for the A-model mirror to D-branes
associated with vector bundles\cite{lsmtwo}. 

While this manuscript was being readied for publication, a
paper\cite{mayr} appeared which in part, overlaps with the contents of
section 6 and 7.

\noindent{\bf Acknowledgments} One of us (S.G.) would like to thank the  
Theory Division, CERN and  two of us (S.G. and T.J.)
the Abdus Salam ICTP for hospitality while this work was being completed.

\end{document}